\documentstyle [aps,epsfig,twocolumn]{revtex}
\draft
\begin{document}
\title{
Stochastic gradient approximation method applied to Hubbard model}

\author{A. Harju~\cite{arimail}}

\address{Laboratory of Physics, Helsinki University of Technology,
P.O. Box 1100, 02015 HUT, Finland}

\date{\today}
\maketitle
\begin{abstract}
The two-dimensional Hubbard model is studied using the variational
quantum Monte Carlo technique with Gutzwiller-type variational wave
functions.  In addition to the simple one-site correlated Gutzwiller
wave function, we use a form with correlation between electrons on
nearest- and next-nearest-neighbor sites.  We show that the stochastic
gradient approximation method suits very well for the optimization of
the free parameters of the variational wave functions.

\end{abstract}

\pacs{PACS numbers: 71.10.Fd, 02.70.Ss, 02.60.Pn}

\section{Introduction}

The one-band Hubbard Hamiltonian is the simplest lattice model for
studying strongly correlated electrons~\cite{Hub}. It is widely
studied, especially in its two-dimensional version, as a model for the
high-$T_c$ superconducting cuprates. Among the various methods used
for studying the Hubbard model, the quantum Monte Carlo (QMC) methods
have shown to be powerful tools, both at zero and finite temperatures.
The studies of the Hubbard model at zero temperature include several
QMC methods, namely, the variational (VMC)~\cite{Yokoyama,Koch}, the
fixed-node diffusion (DMC)~\cite{Koch,Cosentini}, and the constrained
path QMC (CPMC)~\cite{cpmc,cpmc2,cpmc3}.  From these zero temperature
methods, only CPMC does not depend crucially on the variational wave
function.  Actually, the simple free-electron-like wave functions are
rather generally found to be better importance functions for CPMC than
the unrestricted Hartree-Fock wave functions~\cite{cpmc}.  On the
other hand, CPMC is limited to smaller lattice sizes than VMC. It can,
however, give accurate results for system sizes far beyond the limits
of exact diagonalizations~\cite{ED} or the subspace techniques such as
stochastic diagonalization~\cite{SD}.  Also the finite temperature QMC
simulations can be used to obtain ground state properties if the
temperature used is low enough~\cite{hong90}.  This approach is,
however, mainly applicable for half-filled systems due to the problems
related to the negative weights of the configurations, i.e., the
fermion sign problem.

Compared to the other approaches, the VMC methods are very potent for
studying larger system sizes.  The computational cost of the VMC
method is roughly an order of magnitude smaller than of the more
sophisticated QMC methods.  The VMC method is basically free from any
methodological problems such as the fermion sign problem.  There is,
however, one very serious limitation in the VMC approach.  This is
related to the ultimate connection of the method to the many-body wave
function itself. The form of the wave function is an input to the VMC
simulations, and choosing the form requires human creativity and
insight into the problem. Due to this, VMC can hardly be called a
reliable ``black-box'' simulation method. However, the recent
advancements in the optimization of the variational wave
functions~\cite{sga,anad} combined with the enormous increase in
computing power indicate a possibility of having considerably more
flexible VMC wave functions and making the VMC simulations less
``human-biased''.

The crucial part of the VMC simulation is to locate the optimal values
of the free parameters in the many-body wave function.  An efficient
optimization method allows one to have more variational parameters and
thus a more flexible form of the wave function and saves computer
time. In this paper, we show that the optimization method called the
stochastic gradient approximation (SGA)~\cite{sga} suits very well for
the VMC also in the case of lattice Hamiltonians.  We have previously
used it for continuum models, mainly quantum dots~\cite{sga_app}.

The goal in the VMC optimization is to minimize the cost function
\begin{equation}
{\cal F} ( \mbox{\boldmath $\alpha$} ) = \lim_{k \rightarrow \infty}
\frac 1k \sum_{j=1}^k {\cal Q}({\bf R}_j; \mbox{\boldmath
$\alpha$})\ ,
\end{equation}
where configuration ${\bf R}_j$ contains the coordinates of the
simulated particles at $j$th step of a random sequence,
$\mbox{\boldmath $\alpha$} = (\alpha_1,\cdots,\alpha_n)$ represents
the vector of $n$ parameters to be optimized, and ${\cal Q}$ is the
``local'' version of the cost function of a configuration ${\bf R}$
and $\mbox{\boldmath $\alpha$}$.  One can formulate the VMC
optimization problem approximately as a ``standard'' optimization
problem by taking a finite but large $k$ in the cost function above.
One can then use the resulting approximation of the function as a
deterministic function during the optimization. Doing this, one
approximates the whole distribution of ${\bf R}$s by a finite set, and
more importantly, one forgets the $\mbox{\boldmath $\alpha$}$
dependence of ${\bf R}$. To some extent this error can be corrected by
weighting the configurations by factors that depend on the change of
the probability of the configuration in question.

In the SGA optimization method such truncation is not done.  In SGA,
the optimal parameter vector $\mbox{\boldmath$\alpha^*$}$, defined so
that
\begin{equation}
\nabla_{\mbox{\boldmath $\alpha$}} {\cal F} ( \mbox{\boldmath
$\alpha^*$} ) = \lim_{k \rightarrow \infty} \frac 1k \sum_{j=1}^k
\nabla_{\mbox{\boldmath $\alpha$}} {\cal Q}({\bf R}_j; \mbox{\boldmath
$\alpha^*$})  = 0\ ,
\label{opt}
\end{equation}
is found by changing the parameters $\mbox{\boldmath $\alpha$}$ to the
direction of the unbiased stochastic approximation of the negative
gradient, i.e., $-\nabla_{\mbox{\boldmath $\alpha$}} {\cal Q}$.  The
approximate gradient $\nabla_{\mbox{\boldmath $\alpha$}} {\cal Q}({\bf
R}; \mbox{\boldmath $\alpha$}^*)$ is not zero even for the optimal
parameters, but the average of it over configurations ${\bf R}$
vanishes. Due to this, the stochastic simulation should include
damping so that it actually converges. Without damping, the simulation
would end up in oscillating around the optimal parameters. The damping
should, however, be so slow that the simulation is able to reach the
optimal parameters. A more detailed formulation of the SGA method for
VMC is given in Sec.~\ref{method}.

\section{Method}\label{method}

We use the two-dimensional Hubbard model given by the Hamiltonian
\begin{equation}
H = -t \sum_{\langle i,j \rangle \sigma} (c_{i\sigma}^\dagger
c_{j\sigma} + H.c.) + U \sum_{i} n_{i \uparrow} n_{i \downarrow} \ ,
\end{equation}
where $\langle \dots \rangle$ stands for nearest neighbors,
$c_{i\sigma}^\dagger$ ($c_{i\sigma}$) creates (destroys) an electron
with spin $\sigma$ at site $i$ of the square lattice, and $n_{i
\sigma}=c_{i\sigma}^\dagger c_{i\sigma}$. The electrons at the same
site interact with strength $U>0$, and $t$ is the hopping
parameter. We use $t$ as the energy scale, and set $t=1$. The periodic
boundary conditions are used for both space directions.  This choice
of boundary conditions leads to non-interacting single particle states
that are plane-waves $\exp ( i {\bf k \cdot r})$ with energy
$\epsilon_{\bf k}=-2(\cos (k_x)+\cos (k_y))$.

\begin{figure}
\psfig{figure=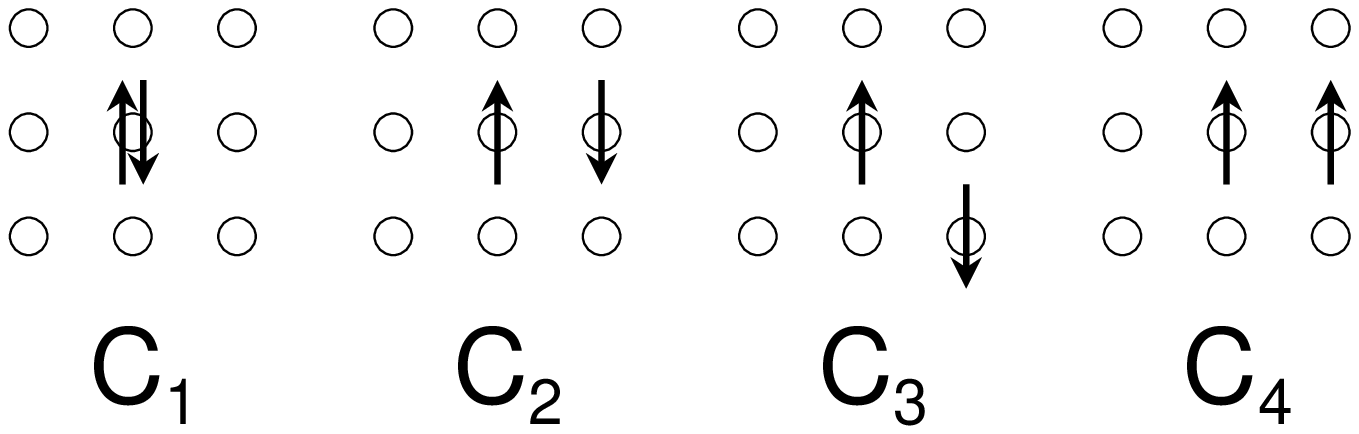,height=0.9\columnwidth,angle=-90}
\caption{Typical configurations measured by the correlation factors in
the Gutzwiller-type wave function. For example, $C_3$ counts the
number of electron pairs that have opposite spin electrons differing
in both coordinates by one.}
\label{Confs}
\end{figure}

A commonly used variational wave function for the Hubbard model is the
Gutzwiller wave function~\cite{Gutz}
\begin{equation}
\Psi(g)=g^D \Psi_0 \ ,
\end{equation}
where $D=\sum_{i} n_{i \uparrow} n_{i \downarrow}$ is the number of
doubly occupied sites, $g$ is the only variational parameter, and
$\Psi_0$ is the many-body wave function for the non-interacting ground
state. In this case, it is made of the plane-waves discussed above,
with the {\bf k} values chosen to minimize the energy. The motivation
for this wave function is that the Gutzwiller factor reduces the
probability to find electrons on the same site and reduces the average
interaction energy. This is done, of course, with a cost of higher
kinetic energy. It is also possible to correlate electrons that are
not on the same site by constructing a Gutzwiller-type wave function
\begin{equation}
\Psi(g)=\sum_{i} g_i^{C_i} \Psi_0 \ ,
\label{genG}
\end{equation}
where the $g_i$'s are parameters and the $C_i$'s measure the number of
nearest (and next nearest) neighbor electrons with various spin
configurations, see Fig~\ref{Confs}. for more details.

We calculate the energy using the standard Metropolis algorithm (see,
e.g., Refs.~\cite{Yokoyama} for more details and Ref.~\cite{Koch} for
a modified algorithm), giving the estimate for the energy as $\bar E =
\langle E_{\rm L}\rangle$, where the local energy is defined as
\begin{equation}
E_{\rm L}(R) = \frac{H \Psi({\bf R})}
{\Psi({\bf R})} \ ,
\end{equation}
and the average in the calculation of $\bar E$ is over configurations
${\bf R}$ distributed according to probability distribution $\propto
|\Psi({\bf R})|^2$.
The standard deviation of the local energy is given by
\begin{equation}
\sigma=\sqrt{\langle (E_{\rm L} - \bar E)^2 \rangle} \ .
\end{equation}
One can also define a modified version of $\sigma$, namely,
\begin{equation}
\tilde \sigma=\sqrt{\langle (E_{\rm L} - E_{\rm T})^2 \rangle} =
\sqrt{\sigma^2+(\bar E-E_{\rm T})^2} \ ,
\end{equation}
which gives the root-mean-square distance of the local energy from the
target energy $E_{\rm T}$. The deviation $\tilde \sigma$ is
particularly useful in the optimization of the wave function.

The SGA optimization method involves stochastic simulation in two
spaces: the configuration and the parameter space. These spaces are
coupled via the parameter vector.  In the configuration space, a set
of $m$ configurations $\{ {\bf R}_j \}$ is sampled from a distribution
$|\Psi(\mbox{\boldmath $\alpha$})|^2$, where $\mbox{\boldmath
$\alpha$}$ is the current parameter vector. When the parameters are
changed, the set of configurations follow this change because the new
sampling distribution depends on the new parameters.  In practice, the
set of configurations is found by the Metropolis algorithm.  In the
parameter space, the parameters at iteration $i+1$ are obtained from
the previous ones by the formula:
\begin{equation}
\mbox{\boldmath $\alpha$}_{i+1}=\mbox{\boldmath $\alpha$}_i-\gamma_i
\nabla_{\mbox{\boldmath $\alpha$}} {\cal Q}_i \ ,
\label{SGA_formula}
\end{equation}
where $\gamma_i$ is a scaling factor of the step length. The scaling
factor has an important role in averaging out the fluctuations in the
approximate gradient, ensuring the convergence. On the other hand, too
small a value of $\gamma$ would damp the simulation too much.  These
rules can be formulated mathematically as:
\begin{equation}
\sum_{i=1}^{\infty} \gamma_i^2 \ < \ \sum_{i=1}^{\infty} \gamma_i \ =
\ \infty \;.
\end{equation}
There is a simple interpretation for these conditions. The sum of
$\gamma^2$ should be finite to dissipate the cumulative error given by
the noise in the approximate gradient and the sum of $\gamma$ should
diverge, because otherwise the maximum distance from the initial
parameters would be limited. If one uses a formula $\gamma_i\propto
i^{-\beta}$, one should have $\frac 12 < \beta \le 1$.  The choice of
$\beta=1$, which is the maximally damped case, leads to a formula
similar to the recursive calculation of a mean:
$\bar{x}_i=\bar{x}_{i-1}-\frac 1i(\bar{x}_{i-1}-x_i)$.

In Eq.~(\ref{SGA_formula}), the approximate gradient is calculated
using the set of $m$ configurations. There are several possibilities
for the cost function ${\cal Q}$. For the energy minimization, the
cost function is simply the mean of the local energies over the set of
configurations:
\begin{equation}
{\cal Q}=\langle E_{\rm L} \rangle= \frac 1m \sum_{j=1}^m E_{\rm
L}({\bf R}_j) \ ,
\end{equation}
and for the variance minimization one has ${\cal Q}=1/m \sum_{j=1}^m (
E_{\rm L}({\bf R}_j) -\langle E_{\rm L} \rangle)^2$.  In the variance
minimization, one can also use a target energy $E_{\rm T}$ instead of
$\langle E_{\rm L} \rangle$.  This leads to the minimization of the
function $\tilde \sigma$ above.

In the original implementation of SGA as presented in Ref.~\cite{sga},
the gradient in Eq.~(\ref{SGA_formula}) was the most difficult task
and was calculated by using a finite-difference formulation.  In this
case, one should note that the configurations are distributed
according to $|\Psi(\mbox{\boldmath $\alpha$})|^2$ and {\em not}
according to $|\Psi(\mbox{\boldmath $\alpha$}\pm \Delta)|^2$, where
$\Delta$ represents a small change.  Thus in the energy minimization,
the finite-difference points are weighted as
\begin{equation}
\frac {1}{\tilde{m}} \sum_{j=1}^m w_j {E}_{\rm L} \left[{\bf
R}_{j}; \mbox{\boldmath $\alpha$}\pm \Delta \right] \ ,
\label{fd}
\end{equation}
where the points ${\bf R}_{j}$ are distributed according to
$|\Psi(\mbox{\boldmath $\alpha$})|^2$, $\tilde{m}=\sum w_j$, and
$w_j=\left| \Psi({\bf R}_{j};\mbox{\boldmath $\alpha$}\pm \Delta) /
\Psi({\bf R}_{j};\mbox{\boldmath $\alpha$})\right|^2$.  The `weights'
$w_j$ of the local functions are very close to unity, because $\Delta$
is only a small change.

It is, however, possible to calculate the gradient also analytically.
In Ref.~\cite{anad}, Lin {\it et al.} have shown that in the case of
real wave function and energy minimization, the derivative of the
energy $E$ with respect to a variational parameter $\alpha_i$ is simply
\begin{equation}
\frac{\partial E}{\partial \alpha_i}=2 \left\{ \left\langle E_{\rm L} \times
\frac{\partial \ln \Psi}{\partial \alpha_i} \right\rangle -E \times
\left\langle \frac{\partial \ln \Psi}{\partial \alpha_i} \right\rangle
\right\} \ ,
\label{anaa}
\end{equation}
where the average $\langle \dots \rangle$ is over the whole Metropolis
simulation~\cite{anad}. One can implement this simple formula also for
the SGA algorithm, with the small modification that the average is
taken over only the current set of $m$ configurations. One should also
note that the finite-difference formulation required calculation of
the local energy with several parameters, whereas only a single
evaluation is needed (for each configuration) if the analytic formula
is used.

In this work, we have used the both schemes discussed above for the
calculation of the gradient.

\section{Results}

\subsection{Single Gutzwiller parameter}

\begin{figure}
\psfig{figure=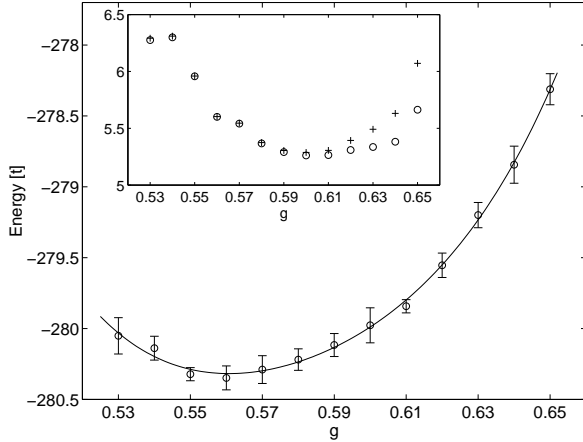,width=0.9\columnwidth}
\caption{Energy as a function of the Gutzwiller parameter $g$. The
solid line shows a fourth-order polynomial fit. The inset shows the
deviations of the local energy, defined as $\sigma$ and $\tilde\sigma$
in the text, marked with '$\circ$' and '$+$', respectively. In
$\tilde\sigma$, $E_{\rm T}=-280.5$ has been used.}
\label{ES_g}
\end{figure}

First, we will consider the wave function with correlation only
between electrons on the same site, and thus only one free parameter
$g$. In Fig.~\ref{ES_g}, the energy and the deviations of the local
energy $\sigma$ and $\tilde\sigma$ are plotted as a function of the
Gutzwiller parameter $g$. The system consists of $101+101$ electrons
on a $16\times16$ lattice with $U=4$. One can see that the minimum of
the total energy is located at $g\approx 0.56$, and the minima of
$\sigma$ and $\tilde\sigma$ are at $g\approx 0.6$. The energy as a
function of $g$ is in a good agreement with Fig.~2. of
Ref.~\cite{Koch}.  The predicted value of $g$ resulting from the
minimization of energy in Ref.~\cite{Koch} is slightly higher, around
$g\approx 0.566$. The statistical error in the energies is too large
in order to locate the optimal value of $g$ with accuracy better than
$0.01$.

\begin{figure}
\psfig{figure=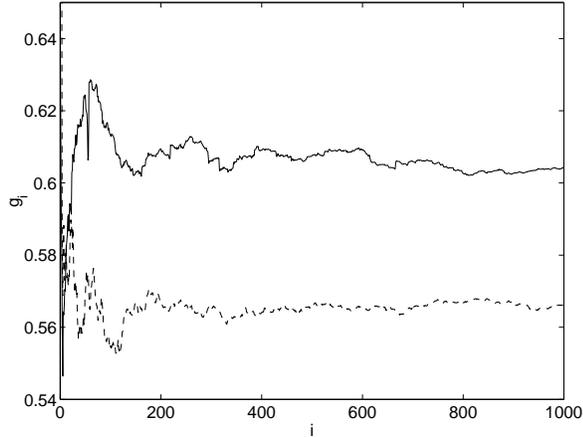,width=0.9\columnwidth}
\caption{The Gutzwiller parameter $g_i$ for the optimization step $i$.
The solid (dashed) line corresponds to the variance (energy)
minimization, respectively. Both simulations are started from
$g_0=0.65$.}
\label{g_i}
\end{figure}

Next, the SGA optimization method is used to locate the optimal values
of $g$, with energy and $\tilde\sigma$ as the cost functions. In
Fig.~\ref{g_i}, the Gutzwiller parameter $g_i$ is plotted as a
function of the SGA optimization step $i$ for both energy and variance
minimizations for the same system parameters as above. We have
calculated the gradient using the original, finite-difference
formulation. The simulations converge to $g\approx 0.566$ in the case
of energy minimization, and to $g\approx 0.603$ in the optimization of
$\tilde \sigma$ with the target energy $E_{\rm T}=-280.5$. Both
parameter values are in a good agreement with the independent
simulations presented above, and with the estimate of the optimal
value of Ref.~\cite{Koch}. One can also see that around 1000 steps are
enough to estimate the optimal parameter values with a reasonable
accuracy. The computational task of this is smaller than of one
independent simulation presented in Fig.~\ref{ES_g} for a single value
of $g$.  On the other hand, it seems that the accuracy of the optimal
parameters found by SGA is one order of magnitude higher than in using
the polynomial fit to the independent points.  We have performed
several simulations starting from different values of $g_0$. In every
case, the performance of SGA is similar to those shown in
Fig.~\ref{g_i}.

\begin{figure}
\psfig{figure=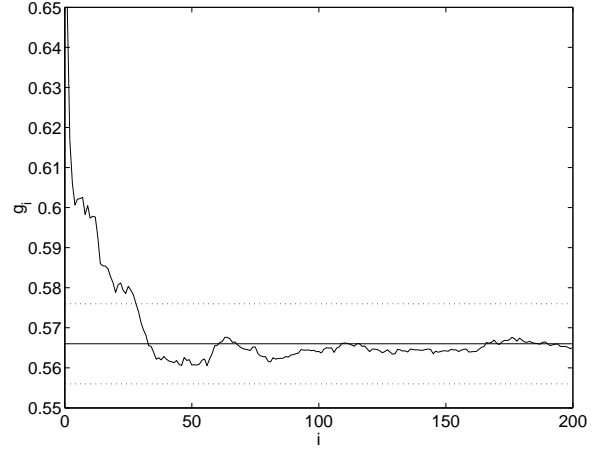,width=0.9\columnwidth}
\caption{The Gutzwiller parameter $g_i$ for the first 200 optimization
steps $i$ using the analytic calculation of the gradient.  Simulation
is started from $g_0=0.65$. The solid line show the optimal parameter
value, and the dotted ones show a range of parameter value that gives
results with energy error within the statistical uncertainty of the
Fig.~\ref{ES_g}. The fluctuation of the parameter value is much
smaller than in Fig.~\ref{g_i}.}
\label{g_in}
\end{figure}

Fig.~\ref{g_in} shows the first 200 values of the Gutzwiller parameter
$g_i$ for the simulation using the analytic derivative of
Eq.~(\ref{anaa}). This simulation converges also very accurately to
the optimal value found above. The most important feature are that the
fluctuations in the value of $g$ during the optimization process is
much smaller than in the simulation where the finite-difference
formula was used. This leads to a faster convergence.

To summarize, the performance of the SGA method has been found very
satisfactory in finding the optimal parameter of the simple
single-parameter Gutzwiller wave function. This is especially true in
the case where the gradient is calculated analytically.

\subsection{Generalized Gutzwiller wave function}

Next, we will consider the generalized Gutzwiller wave function,
defined in Eq~(\ref{genG}). We consider only the generalization with 4
parameters related to typical configurations shown in
Fig~\ref{Confs}. The aim of these parameters is to capture partially
the missing correlations.  The generalization made to the Gutzwiller
wave function is bosonic in the sense that it does not change the
nodal structure of the one-parameter Gutzwiller wave function. The
energy gain of this extra correlation could directly be compared to
the energy of the fixed-node DMC simulations of Ref.~\cite{Koch} as
the nodal structure used there is also given by the simple Gutzwiller
wave function. This DMC energy is lower than the corresponding VMC
energy by $\approx 2.5$ units.  One should note that in the lattice
formulation of DMC, unlike in the continuum formulation, the energy
depends also on the bosonic correlation factor which does not change
the nodes. This dependence is much smaller than the dependence of the
VMC energy on $g$.  There are also correlations that change the nodal
structure, and the estimate for the importance of these can be
estimated from the CPMC energy which is lower than the VMC energy by
more than 6 units~\cite{cpmc}.

We have optimized the 4 free parameters of the generalized Gutzwiller
wave function using the SGA and the analytic formulation of the
gradient. The latter choice is due to better results in the
one-parameter case. One should note that setting the parameter $g_1=g$
and $g_{2}=g_{3}=g_{4}=1$ one obtains the simple Gutzwiller wave
function. We have again studied $101+101$ electrons on a $16\times16$
lattice with $U=4$ as above.

The optimization of the energy leads to a reduced on-site correlation
factor of $g_1\approx 0.51$. The other parameters are $g_2\approx
0.92$ and $g_3 \approx g_4 \approx 0.98$. The smaller value of $g_1$
does not cost as much kinetic energy in this case as in the case of
a simple Gutzwiller wave function, because the values of $g_2$ and $g_3$
are also smaller than one. It is also interesting to compare the
optimal value of $g_1$ found here with the optimal $g\approx 0.52$ of
the simple Gutzwiller wave function determined from the $g$ dependence
of the DMC energy~\cite{Koch}. The optimization converges again in few
hundred steps, in similar fashion as shown in Fig.~\ref{g_in}.

The energy calculated with the optimal parameters is presented in the
Table~\ref{t2} with corresponding one-parameter VMC, DMC, and CPMC
results. One can see that the generalization of the Gutzwiller wave
function is able to lower the energy by 1.1 units, which is less than
half of the difference to the DMC energy. The CPMC energy is still
around 5 units lower in energy.

There are still important ingredients missing from the variational
wave function. Possible extensions are three- and higher-particle
correlation factors, modified single-particle states, and
multiple-determinant wave functions. The good performance of the SGA
method combined with the simple calculation of the energy gradient
could be extremely useful in the studies exploring these directions.

\section{Conclusions}

We have shown that the SGA optimization method finds the optimal
values of the Gutzwiller wave function parameter in a reliable and
efficient fashion. The computational cost of the optimization process
is comparable to a single calculation of the expectation values with a
fixed parameter value.  The good performance is very important
particularly if a variational wave function with several parameters is
used.  Due to this, SGA is very useful in finding more accurate wave
functions for the Hubbard model.

\acknowledgments{We would like to thank E. Koch, C. Dahnken and
A.C. Cosentini for helpful discussions and R.M. Nieminen for reading
the manuscript.}

\begin{table}
\caption{Energy of $101+101$ electrons on a $16\times16$ lattice with
$U=4$ for different methods. Numbers in the parenthesis are
statistical errors in energy. The number of VMC parameters is also
given. The DMC energy is from Ref.~\protect\cite{Koch} and CPMC from
Ref.~\protect\cite{cpmc}.}
\begin{tabular}{cc}
Method   & Energy \\ \hline
VMC 1 & -280.3(1) \\
VMC 4 & -281.4(1) \\
DMC & -283.0(5)\\
CPMC & -286.55(8)
\end{tabular}
\label{t2}
\end{table}

\end{document}